\documentstyle{article}

\title{Quantum Mechanics as a Classical Theory III:\\
Epistemology}
\author{L.S.F. Olavo\\
Departamento de Fisica, Universidade de Brasilia,\\
70910-900 - Brasilia - D.F. - Brazil}

\begin{document}

\maketitle
\begin{abstract}
The two previous papers developed quantum mechanical formalism from
classical mechanics and two additional postulates. In the first paper it was
also shown that the uncertainty relations possess no ontological validity
and only reflect the formalism's limitations. In this paper, a Realist
Interpretation of quantum mechanics based on these results is elaborated and
compared to the Copenhagen Interpretation. We demonstrate that von Neumann's
proof of the impossibility of a hidden variable theory is not correct,
independently of Bell's argumentation. A local hidden variable theory is
found for non-relativistic quantum mechanics, which is nothing else than
newtonian mechanics itself. We prove that Bell's theorem does not imply in a
non-locality of quantum mechanics, and also demonstrate that Bohm's theory
cannot be considered a true hidden variable theory.
\end{abstract}

\section{Introduction}

The first two papers developed quantum mechanics within the fundamental
principles of it's mathematical formalism. For this, we use no more than
classical mechanics and two rather natural postulates that do not modify its
classical character.

We demonstrated in the first paper that the fundamental equations are those
that involve the density function. We also demonstrated that the uncertainty
relations are not valid for these equations and are, for this reason, an
indication of the formalism's limitation of using equations for the
probability amplitudes. The uncertainty relations form the basis of the
Copenhagen Interpretation\cite{1,2} and the contestation of their
ontological character has profound implications for quantum mechanics
epistemology.

A model in which the introduction of an observing system in a specific
quantum problem is considered was constructed. This observing system has
profound differences from those introduced by quantum mechanics' diverse
measurement theories\cite{3}-\cite{9}.

In the present paper, we will develop a Realist Interpretation for quantum
mechanics based on the results obtained in the first and second papers
(hereafter abbreviated by (I) and (II)).

In the second section, we will quickly review the fundamental ideas of the
Copenhagen Interpretation.

In the third section, we will establish a Realist Interpretation for quantum
mechanics based on the results of this series' first two papers\cite{10}.
Certain points of the Copenhagen interpretation presented in the previous
section will also be criticized.

The fourth section will show that von Neumann's demonstration of the
impossibility of a hidden variable theory is not correct, independently of
Bell's argument\cite{11}, which we prove inadequate. A local variable theory
is then found for the non-relativistic problem which is nothing more than
newtonian mechanics.

In the fifth section, we demonstrate that Bell's theorem does not imply in a
non locality of quantum mechanics.

In the sixth section, we reinterpret Bohm's theory\cite{7,12} and show that
it cannot be considered a true hidden variable theory.

In the last section, we make our final conclusions for this series of papers.

\section{The Copenhagen Interpretation}

We will present the main ideas of the Copenhagen Interpretation through
a\-xi\-oms\cite{13}. We are not interested in the formalism, but in the
interpretation which accompanies it. For this reason, the formal apparatus
will not be developed.

Let us first define the meaning of quantum state:

\begin{description}
\item[(def1)]  The state of a physical system is represented by $\phi $, a
function of {\it s} coordinates. This function does not necessarily
represent any distribution of objects within space, because the {\it s}
variables that index it possess no intuitive association with objects. It is
not defined in terms of observables but only as a function within
configuration space. These {\it s} variables represent the system's degrees
of freedom.
\end{description}

The Copenhagen Interpretation's axioms are\cite{13}:

\begin{description}
\item[(CAx1)]  The function $\phi $, which, in general, can be complex
should be square integrable.

\item[(CAx2)]  The function $\phi $ should be single valued.

\item[(CAx3)]  For every observable $p$ there is a single operator $P$
acting upon the state function. In particular, we obtain Schroedinger's
equation (for the amplitudes) substituting, in the hamiltonian function, the
variables $q$ and $p$ by their respective operators.

\item[(CAx4)]  The only possible values that can be observed when a
measurement is made over an observable $p$ are the eigenvalues of the
following equation
\begin{equation}
\label{(1)}P\psi _\lambda =p_\lambda \psi _\lambda ,
\end{equation}
where $\psi _\lambda $ satisfies axioms (CAx1) and (CAx2).

\item[(CAx5)]  When a system is in a certain state $\phi $, the expected
value of a series of measurements of the observable $p$ is
\begin{equation}
\label{(2)}\overline{p}=\frac{\int \phi ^{*}P\phi d\tau }{\int \phi ^{*}\phi
d\tau },
\end{equation}
where $P$ is the operator which corresponds to $p$.
\end{description}

All of quantum mechanical formalism can be obtained from these five axioms.
We will not do this here. Nevertheless, one particular result which
interests us will be presented, without demonstration, as the following
theorem:

\begin{description}
\item[(CT1)]  If $Q$ and $P$ are canonically conjugated operators, then the
dispersion associated with the simultaneous measurements of their
eigenvalues is given by
\begin{equation}
\label{(3)}\left[ Q,P\right] =i\hbar .
\end{equation}
\end{description}

Even though this result is a theorem, it's interpretation is fundamental for
the Copenhagen Interpretation of quantum mechanics and it is called the
Uncertainty Principle. Thus, according to the Copenhagen Interpretation\cite{%
14}, we have the following interpretative postulates:

\begin{description}
\item[(CP1)]  In a given experiment, it is not possible to measure with
absolute precision both the position and the momentum of a given particle.
The minimum dispersion in such a measurement is given by relation (\ref{(3)}%
).
\end{description}

Because it adequately describes nature's behavior, this postulate possess
ontological {\it status}. Nevertheless, this description is not objective,
but subjective, and entails:

\begin{description}
\item[(CP2)]  Any physical system's attributes, for example, its
trajectories, only come into existence when observed.
\end{description}

and more,

\begin{description}
\item[(CP3)]  An inevitable and uncontrollable interaction occurs between
the measured object and the observer.
\end{description}

The dispersion relation (\ref{(3)}) served as the mathematical element for
Bohr to expose his ideas about the complementarity which he had been
developing since his first contact with the dual wave-particle behavior of
some experiments. Once it is assumed that there can be no space-time
coordination, causality and the notion of a complete classical description
become obsolete\cite{15}. Thus

\begin{description}
\item[(CP4)]  Combining a classical observational system with a quantum
system, one can only measure complementary values and, to express them in
classical terms, the system must also be described in terms of classical
figures which are also complementary.
\end{description}

Finally, to establish the state function's referent, it is postulated that

\begin{description}
\item[(CP5)]  The wave function expresses our knowledge of events.
\end{description}

In relation to the influence of the act of observing on measurement, we can
say that

\begin{description}
\item[(CP6)]  An unforeseeable and discontinuous reduction of the state
vector, formally represented by a projection operation occurs during the act
of measurement.
\end{description}

These are, shortly put, the main ideas which form the Copenhagen
Interpretation. In the next section we will criticize these ideas through
the results obtained in this series of papers.

\section{The Realist Interpretation}

In this section we will demonstrate how a realist view can be made
compatible with quantum mechanics' formalism as developed in the two
previous papers.

First, it is important to note that the observer, massively discussed in the
various quantum mechanical interpretations, does not play any role in it's
formalism. When any problem is to be solved, such as to find the energy
levels of an atomic system for example, the possible interactions between
this system and the external world are never formally taken into account.
This can be seen in the present formalism through the derivations of the
quantum equations from the Liouville's equation for a closed system. The
introduction of the observer into the orthodox epistemology is not only {\it %
ad hoc}, it is also incompatible with quantum mechanic's main postulate,
which is Schroedinger's equation, since a discontinuous reduction of the
state vector, which does not satisfy this equation\cite{8}, must be
considered when the observer is taken into account.

Here we wish to differentiate statements such as ''the value of the property
$P$ of the physical object $y$ is equal to $x$'' from others such as: ''the
observer $z$ found the value {\it y} for the property $P$ of the physical
object $x$ using the measurement technique $t$ and the sequence of
operations $o${\it ''}. In fact, while the first proposition can be
mathematically represented as $P(x)=y$, the second one should be given as $%
P^{\prime }\left( x,z,t,o\right) $, so that $P$ and $P^{\prime }$ possess
radically different referents\cite{16}.

We can give an example of these concepts using the treatment given to an
external observer system as applied to Young's interference experiment in
(I). In the usual treatment given through Schroedinger's equation, we have a
state function which will depend only on the variables $F\left( {\bf x}_1,%
{\bf p}_1;t\right) $ associated to the system's internal degrees of freedom.
In the treatment given in (I), we should consider not only $F\left( {\bf x}%
_1,{\bf p}_1;t\right) $, but also the observing system's probability density
function $F\left( {\bf x}_2,{\bf p}_2;t\right) $. Any variation of the
pragmatic parameters which define this distribution can alter the
probability distribution which we intend to calculate.

In the measurement theory associated to the Copenhagen Interpretation, the
observer's conscience must be postulated to avoid an infinite regression of
reductions in the state vector\cite{3,4,5,8}. This infinite regression can,
nevertheless, be easily understood through the present theory's point of
view. As was observed by Everett\cite{6}, the question about the infinite
regression of reductions in the state vector is intimately linked to
considerations of open and closed systems. When studying the Young's
interference experiment, we considered {\it system2} as an external observer
and take into account its interaction with {\it system1} through the concept
of scattering cross-section. If the {\it system1} and {\it system2} are to
be considered as part of a greater system, then their interaction will be
taken into account exactly and the equation representing the probability
distribution of the whole system will be $F\left( 1,2\right) $, which is, in
general, very different from the uncoupled distributions $F_1\left( 1\right)
$ and $F_2\left( 2\right) $ used to represent the open system.

In fact, the equation obtained in (I) for the probability density function
of this experiment can no longer be transformed, through the Wigner-Moyal
Infinitesimal Transformation, into the Schroedinger equation for the
probability amplitude. Thus, the observed system's wave-like behavior of the
{\it ensemble}'s statistics no longer needs to manifest itself. It will
manifest itself depending on the intensity of the perturbation introduced by
the observer system.

Of course, this is not an uncontrollable interaction between the measuring
apparatus and the observing system. In fact, this idea emerged from Bohr's
quantum postulate which saw the finite character of Plank's constant,
interpreted by him as associated with the minimal quantity of energy a
system can emit or absorb, as representing the impossibility of analysis.
Note that, as was shown in (I), Plank's constant can't be used as a
characteristic entity of a quantum system since that which is conventionally
called the classical limit is independent of it.

Schroedinger's equation for the probability amplitudes represents systems in
equilibrium. The absorption or emission of energy in these systems is done
in such a manner that the system passes from one equilibrium state to
another. It is this that we call the quantum jump. Of course, the
non-equilibrium states are not subjected to such considerations (two-photon
absorption is one example of such a situation). Observing the equation in
which the external observing system is introduced through a Boltzmann
equation, it can be seen that the quantum equation obtained {\it does not},
in general, have eigenvalues. Therefore, it is neither associated with
quantized values of energy nor equilibrium situations: it does not represent
quantum jumps. {\it Natura non facit saltus}. It should be also mentioned
that, in the second paper, it was shown that strong gravitational fields
will, in general, deny equilibrium situations and no quantization is
expected. So, quantization cannot be considered as a fundamental
manifestation of Nature.

We note also that in this interpretation the problem of considering the
frontier between the classical system, usually the external observer, and
the observed quantum system, does not occur. In the same manner, there is a
symmetry in treatment, because the observing system can be considered an
observed system and {\it vice-versa}, this being an important property if we
desire to make relativistic generalizations of our arguments.

It becomes clear from the discussion above that a general theory of
measurement, as was proposed by von Neumann\cite{3}, has few chances of
success since, to admit such a theory, supposing that the description of the
observer should appear in the equations, we must also admit that there is a
way to unify, under the same formal apparatus, the innumerable existing
observational techniques.

Having clarified the issue of the observer in quantum mechanics, we pass to
a criticism of the postulates presented in the previous section:

\begin{description}
\item[(CP1)]  This postulate, as has been observed, has no ontological
validity. Instead of determining a limitation to the classical concepts of
space-time coordination, it determines a limitation of the quantum formalism
given by Schroedinger's Second Equation for the probability amplitudes.

\item[(CP2)]  Without the uncertainty relations, this postulate suggests a
confusion between the concepts of having a known value and having a define
value. Unlike what is stated by the Copenhagen Interpretation, the variables
which index the density function do refer to the state of the {\it ensemble's%
} constituent components. Even if we do not know these values in a specific
experimental arrangement, and for this reason treat them statistically, we
assume that these values are well defined. In fact, in the limit of
dispersion free {\it ensembles}, we reobtain newtonian trajectories.

\item[(CP3)]  We have seen that this postulate is associated to the fact
that the Schroe\-dinger equation for the probability amplitudes describes
systems in an equilibrium situation and is therefore incapable, in this
format, of giving information to the inter-phenomenon occurrences. In any
manner, when we introduce the external observing system, we perceive that
its interaction with the observed system is classical and controllable
(deterministic), even if it is considered unknown and treated statistically.

\item[(CP4)]  Once we have denied the ontological validity of the dispersion
relations and assumed the figure of space-time coordination, we can no
longer accept this postulate.

\item[(CP5)]  There are no external psycho-physical observers. The density
function supplies an objective description of the physical world and does
not possess any relation with mental activities.

\item[(CP6)]  See the criticism to postulate (CP3) and the discussion before
it.
\end{description}

In relation to the wave-particle duality, it has been demonstrated that all
the results in which the systems present wave-like behavior can be explained
through the quantized interaction which occurs between them\cite{17,18}
through the use of the particle picture.

We are, at this point, ready to accept the definitions of reality and
completion given by Einstein, Podolsky and Rosen\cite{19}. Yet there is,
still, one problem. It has been recently proven that realist theories,
called hidden variable theories, should possess non local behavior. Even if
this behavior can be accepted by a realist theory, it contradicts the
relativistic theories from which we derived quantum formalism presented in
(II)\ (remember that, since there is no conscious observer in this theory,
no distinction is made between signals and information as is usually done by
those who wants to suppress the above cited contradiction). Such behavior is
therefore, inconceivable within such formalism.

We will demonstrate, in the next sections, that quantum mechanics is a local
theory.

\section{Hidden Variables}

In the previous section it was seen that, in sight of the results obtained
in (I) and (II), quantum mechanics admits an epistemology completely
different from that accepted as orthodox since the Solvay Congress in 1927.
It has also been shown that this theory, of statistical character, is built
upon a totally deterministic and local theory, which is no more than
newtonian mechanics (in a non-relativistic approximation). In this manner,
newtonian mechanics can be considered {\it the} quantum mechanic's hidden
variable theory.

If this is true, we should demonstrate that von Neumann's proof\cite{3} on
the impossibility of such a theory is mistaken.

In this section and the following, we will demonstrate that newtonian
mechanics can be considered the hidden variable theory for quantum mechanics
and that von Neumann's argument about the impossibility of a hidden variable
theory is incorrect, independently of Bell's argumentation\cite{11}.

Von Neumann's axioms are

\begin{description}
\item[(A1)]  If an observable is represented by the operator $R$, then this
observable's function $f$ is represented by $f(R)$.

\item[(A2)]  The sum of many observables $R,S,..$ is represented by the
operator $R+S+..$ whether they commute or not.

\item[(A3)]  The correspondence between operators and observables is one to
one.

\item[(A4)]  If the observable $R$ is non negative, then its mean value $<R>$
is also non negative.

\item[(A5)]  For arbitrary observables $R,S,..$ and arbitrary real numbers $%
a,b,..$ we have
\begin{equation}
\label{(4)}\left\langle aR+bS+...\right\rangle =a\left\langle R\right\rangle
+b\left\langle S\right\rangle +...,
\end{equation}
for all the possible {\it ensembles} for which the mean values can be
calculated.
\end{description}

{}From these axioms, von Neumann obtains the density operator $\rho $ by
construction, together with it's properties. Among these properties is it's
use for the calculation of the observable's mean values
\begin{equation}
\label{(5)}Tr\left( \rho R\right) =\left\langle R\right\rangle .
\end{equation}

Yet, von Neumann argues that, for any hidden variable theory, we should have
dispersion free states for which
\begin{equation}
\label{(6)}\left\langle R^2\right\rangle -\left\langle R\right\rangle ^2=0.
\end{equation}

Using the result (\ref{(5)}) in (\ref{(6)}) for the observable
\begin{equation}
\label{(7)}R=\left| \phi \right\rangle \left\langle \phi \right| ,
\end{equation}
we reach
\begin{equation}
\label{(8)}\left\langle \phi \right| \rho \left| \phi \right\rangle
=\left\langle \phi \right| \rho \left| \phi \right\rangle ^2,
\end{equation}
for every amplitude $\left| \phi \right\rangle $.

In this case, von Neumann concludes that $\rho =0$ or $\rho =1$. The first
hypothesis has no physical meaning and the second does not imply a
dispersion free state for vector spaces with more than one dimension. In
fact, in case of a space of dimension $d$ we get
\begin{equation}
\label{(9)}Tr\left( \rho \right) =Tr\left( 1\right) =d,
\end{equation}
Thus, the expression to the left of (\ref{(6)}) becomes
\begin{equation}
\label{(10)}\left\langle R^2\right\rangle -2\left\langle R\right\rangle
^2+\left\langle R\right\rangle ^2\left\langle 1\right\rangle ,
\end{equation}
which is not equal to zero if $d\geq 2$. Thus, it is not possible to
construct a hidden variable theory compatible with quantum mechanics.

With the advent, in 1952, of a supposedly hidden variable theory\cite{12}
totally compatible with quantum mechanics, it became necessary to
demonstrate that von Neumann's demonstration contained some mistake. To do
this, Bell defended that the fifth axiom was responsible for this
inconsistency; his argument was the following\cite{20}:

\begin{description}
\item[B-arg.:]  ''At first sight the required additivity of expectation
values seems very reasonable, and it is rather the non-additivity of allowed
values (eigenvalues) which requires explanation. Of course the explanation
is well known: A measurement of a sum of non commuting observables cannot be
made by combining trivially the results of separate observations on the two
terms - it requires a quite distinct experiment. For example the measurement
of $\sigma _x$ for a magnetic particle might be made with a suitably
oriented Stern-Gerlach magnet. The measurement of $\sigma _y$ would require
a different orientation, and of $\left( \sigma _x+\sigma _y\right) $a third
and different orientation. But this explanation of the non-additivity of
allowed values also established the non triviality of the additivity of
expectation values. The latter is quite a peculiar property of quantum
mechanical states, not to be expected {\it a priori}. There is no reason to
demand it individually of the hypothetical dispersion-free states, whose
function it is to reproduce the {\it measurable} peculiarities of quantum
mechanics {\it when averaged over}.''
\end{description}

Obviously this argument cannot be accepted by the present theory. For the
present theory, measurements are done in quantum mechanics exactly as in
classical statistical mechanics and therefore possess the same
characteristics. The argument says, beyond this, that the non-triviality of
measures is due to the non-commutativity of the observables which one is
measuring. Since we have proven that this non-commutativity has no
ontological validity, we should reject this argument.

It can, nevertheless, be shown that von Neumann's demonstration is
incorrect. Let us now pass on to this demonstration:

The {\it ensembles'} state is determined, in phase space, by the density
functions $F\left( {\bf x},{\bf p};t\right) $. For a realist theory, a
dispersion free {\it ensemble} constituted by $N$ particle systems is
represented by the product
\begin{equation}
\label{(11)}F\left( {\bf x}_1..{\bf x}_N;{\bf p}_1..{\bf p}_N;t\right)
=\prod_{i=1}^N\delta \left( {\bf x}_i-{\bf x}_i^0\left( t\right) \right)
\delta \left( {\bf p}_i-{\bf p}_i^0\left( t\right) \right) ,
\end{equation}
where each pair of Dirac's delta functions determines the trajectory -
deterministic, causal and given by newtonian mechanics - of one of the
components of the {\it ensemble's} constituent systems.

Using the Wigner-Moyal Infinitesimal Transformation we obtain the density
function
\begin{equation}
\label{(12)}\rho \left( {\bf x}_1,\Delta {\bf x}_1;..;{\bf x}_N,\Delta {\bf x%
}_N;t\right) =\prod_{i=1}^N\delta \left( {\bf x}_i-{\bf x}_i^0\left(
t\right) \right) \exp \left[ \frac i\hbar {\bf p}_i^0\left( t\right) \cdot
\Delta {\bf x}_i\right] ,
\end{equation}
using $\Delta x$ in order not to confuse them with the Dirac's delta
functions. Taking the limit $\Delta x\rightarrow 0$, we obtain the density
\begin{equation}
\label{(13)}\rho \left( {\bf x}_1{\bf ,..,x}_N;t\right) =\prod_{i=1}^N\delta
\left( {\bf x}_i-{\bf x}_i^0\left( t\right) \right) ,
\end{equation}
as expected. Integrating this expression we find
\begin{equation}
\label{(14)}\int \rho \left( {\bf x}_1{\bf ,..,x}_N;t\right) d{\bf x}_1{\bf %
..x}_N=1,
\end{equation}
which is in clear contradiction with expression (\ref{(9)}). In this manner
we always have expression (\ref{(10)}) equal to zero. It is important to
note that the operations of taking the limit in (\ref{(13)}) and integrating
in (\ref{(14)}) are equivalent to take the trace $Tr\left( \rho \right) $.

We can therefore say that, from expression (\ref{(9)}) and for dispersion
free states, we cannot conclude, as von Neumann did, that $\rho =0$ or $\rho
=1$.

Note that we demonstrated the incorrectness of von Neumann's theorem above
and, simultaneously showed that newtonian mechanics may be the hidden
variable theory behind quantum mechanics. Yet this theory is local. Bell,
extending the argument above, proved through a theorem, that every hidden
variable theory should be non-local. We should, therefore, analyze his
theorem.

\section{Bell's Theorem}

Consider two meters measuring two particles which are the product of a
physical system's dissociation. The results given by these meters are
represented by $A\left( {\bf a},\lambda \right) $ and $B\left( {\bf b}%
,\lambda \right) $, where ${\bf a}$ and ${\bf b}$ are the meter's
orientations and $\lambda $ is a set of hidden variable with a probability
density $\rho \left( \lambda \right) $, which determine the quantum state of
each of the {\it ensemble's} component systems. Writing the results in this
manner we are assuming the locality thesis, since the values measured by a
meter, {\it A} for example, do not depend on the other's configuration (in
this case the ${\bf b}$ orientation).

We now ask if the correlation
\begin{equation}
\label{(15)}P\left( {\bf a},{\bf b}\right) =\int \rho \left( \lambda \right)
A\left( {\bf a},\lambda \right) B\left( {\bf b},\lambda \right) d\lambda ,
\end{equation}
where
\begin{equation}
\label{(16)}\int \rho \left( \lambda \right) d\lambda =1,
\end{equation}
can be equal to the value obtained through the quantum mechanical
calculation. Suppose, for generality
\begin{equation}
\label{(17)}\left| A\left( {\bf a},\lambda \right) \right| \leq 1\quad
;\quad \left| B\left( {\bf b},\lambda \right) \right| \leq 1\quad ;\quad
\left| P\left( {\bf a},{\bf b}\right) \right| \leq 1,
\end{equation}
We obtain, after some algebra that is independent of quantum mechanical
considerations, the following inequality
\begin{equation}
\label{(18)}\left| P\left( {\bf a},{\bf b}\right) -P\left( {\bf a},{\bf b}%
^{\prime }\right) \right| +\left| P\left( {\bf a}^{\prime },{\bf b}^{\prime
}\right) +P\left( {\bf a^{\prime }},{\bf b}\right) \right| \leq 2,
\end{equation}
which is called Bell's inequality and should be obeyed by predictions of
local theories.

{}From here Bell shows that the quantum correlation of spins does not obey
this inequality. Bell's theorem thus states that no local hidden variable
theory can reproduce quantum mechanics' results. His argumentation based on
the experiment proposed by Bohm and Aharonov\cite{21}, is the following:

Take an {\it ensemble} of systems initially in singlet form. The probability
amplitude associated to this {\it ensemble} is given by
\begin{equation}
\label{(19)}\left| \Psi _S\right\rangle =\frac{\left| +\right\rangle \left|
-\right\rangle -\left| -\right\rangle \left| +\right\rangle }{\sqrt{2}}.
\end{equation}

At a given moment, this system dissociates itself into {\it particle1} and
{\it particle2} which are measured by {\it meter1} in direction ${\bf a}$
and {\it meter2} in direction ${\bf b}$. In this case
$$
\left\langle \Psi _S\right| \sigma _a\sigma _b\left| \Psi _S\right\rangle
=\frac 12\cdot \left[ \left\langle +\right| \sigma _a\left| +\right\rangle
\left\langle +\right| \sigma _b\left| +\right\rangle -\left\langle +\right|
\sigma _a\left| -\right\rangle \left\langle -\right| \sigma _b\left|
+\right\rangle -\right.
$$
\begin{equation}
\label{(20)}\cdot \left. -\left\langle -\right| \sigma _a\left|
+\right\rangle \left\langle +\right| \sigma _b\left| -\right\rangle
+\left\langle -\right| \sigma _a\left| -\right\rangle \left\langle +\right|
\sigma _b\left| +\right\rangle \right] ,
\end{equation}
where $\theta _{ab}$ is the angle between ${\bf a}$ and ${\bf b}$.

Now placing ${\bf a}$ in direction ${\bf z}$, we obtain
\begin{equation}
\label{(21)}\left\langle \Psi _S\right| \sigma _a\sigma _b\left| \Psi
_S\right\rangle =-\cos \theta _{ab}.
\end{equation}
This result violates Bell's inequality, when the arrangement
\begin{equation}
\label{(22)}\angle {\bf ab}^{\prime }=2\theta \quad ;\quad \angle {\bf ba}%
^{\prime }=0\quad ;\quad \angle {\bf bb}^{\prime }=\angle {\bf aa}^{\prime
}=\theta ,
\end{equation}
is made along with the rotational symmetry of the calculation made in (\ref
{(21)}). Then Bell is in position to interpret this result as saying that
quantum mechanics has a non-local character.

Let us now suppose that the state of the {\it ensemble} that is to be
measured is prepared in such a way that only one of it's component systems
is measured at a time. To represent this system's state after the separation
we have
\begin{equation}
\label{(23)}\left| \Psi \right\rangle =\left| +\right\rangle \left|
-\right\rangle \quad or\quad \left| \Psi \right\rangle =\left|
-\right\rangle \left| +\right\rangle .
\end{equation}
This must be so since, according to Born's statistical interpretation, each
system has probability one half to be in only one of the states above
mentioned (in fact this question is related to that one about the state
vector representing the state of {\it one system} - Schroendiger's dead and
alive cat - or representing an {\it ensemble} of states; we are clearly
opting for the former interpretation). In this case the correlation is given
by
\begin{equation}
\label{(24)}\left\langle \Psi _S\right| \sigma _a\sigma _b\left| \Psi
_S\right\rangle =\frac 12\left[ \left\langle +\right| \sigma _a\left|
+\right\rangle \left\langle +\right| \sigma _b\left| +\right\rangle
+\left\langle -\right| \sigma _a\left| -\right\rangle \left\langle +\right|
\sigma _b\left| +\right\rangle \right] .
\end{equation}

For this correlation and for the configuration (\ref{(22)}), we obtain the
expressions%
$$
P\left( {\bf a},{\bf b}\right) =-\cos \theta \quad ;\quad P\left( {\bf %
a^{\prime }},{\bf b}\right) =-\cos {}^2\theta
$$
\begin{equation}
\label{(25)}P\left( {\bf a},{\bf b^{\prime }}\right) =-\cos 2\theta \quad
;\quad P\left( {\bf a^{\prime }},{\bf b}\right) =-\cos \theta \cos 2\theta
\end{equation}
which, when substituted in Bell's inequality, give us the expression
\begin{equation}
\label{(26)}\left| \cos 2\theta -\cos \theta \right| +\left| \cos \theta
\right| \left| \cos 2\theta +\cos \theta \right| \leq 2
\end{equation}
which, it can be shown, is always satisfied.

It becomes clear from these two treatments that the difference is caused by
the fact that we have not considered, in (\ref{(24)}), the terms
\begin{equation}
\label{(27)}\left\langle +\right| \sigma _a\left| -\right\rangle
\left\langle -\right| \sigma _b\left| +\right\rangle \quad ;\quad
\left\langle -\right| \sigma _a\left| +\right\rangle \left\langle +\right|
\sigma _b\left| -\right\rangle
\end{equation}
which represent interference effects acting upon only one system.

On the other hand, if we realize an experiment where {\it many systems} are
measured simultaneously, then we expect the terms (\ref{(27)}) to appear,
even if representing a correlation between {\it different systems}. This
correlation, evidently, cannot be used to probe the non-local character of a
theory.

The discussion above shows that quantum mechanics is a local theory because
it satisfies Bells inequality (in fact it remains to prove, by experiments,
if the statistical interpretation adopted for the state vector is correct).

Many experiments were realized to demonstrate the violation of Bell's
ine\-qua\-li\-ty\cite{22}-\cite{30}; nevertheless, all these experiments, as
far as
we could see, perform measurements over various systems at a time, bringing
results which agree perfectly well with quantum mechanic's predictions, as
expected\cite{25}. We here propose that experiments be made in which the
systems are measured {\it one by one} in order to confirm result (\ref{(26)}%
) and validate the statistical interpretation of the state vector proposed.
Some of these experiments have already been done. Indeed, experiments on
Interrupted Florescence\cite{31}-\cite{34} show that the hypotesis made in
expression (\ref{(23)}) about the correct appearance of the probability
amplitudes for a single system is adequate.

\section{Bohm's Theory}

Bohm's theory has been considered an authentic hidden variable theory ever
since it was published in 1952. It was this theory that resurrected the
discussion about the possibility of hidden variable theories.

In this section we will give an argument in order to show that this theory
may not be considered a true hidden variable theory. For this, we will
present it shortly.

{}From Schroedinger's second equation, and writing the amplitude of
probability as
\begin{equation}
\label{(28)}\Psi =R\left( x\right) \exp \left[ iS\left( x\right) /\hbar
\right] ,
\end{equation}
where $R\left( x\right) $ and $S\left( x\right) $ are real functions, Bohm
obtains, equating the real and imaginary terms to zero, the following
equations
\begin{equation}
\label{(29)}\frac{\partial S}{\partial t}+\frac{\left( \nabla S\right) ^2}%
m+V+Q=0,
\end{equation}
\begin{equation}
\label{(30)}\frac{\partial P}{\partial t}+\nabla \left( P\frac{\nabla S}%
m\right) =0,
\end{equation}
where $V\left( x\right) $ is the classical potential (we have done this in
reverse order in (I) and (II)). He calls $Q\left( x\right) $ the quantum
potential, defined as
\begin{equation}
\label{(31)}Q=-\frac{\hbar ^2}{2mR}\nabla ^2R
\end{equation}
and
\begin{equation}
\label{(32)}P\left( x;t\right) =R\left( x;t\right) ^2=\Psi ^{*}\left(
x;t\right) \Psi \left( x;t\right) .
\end{equation}

Using Hamilton-Jacobi formalism, the equation
\begin{equation}
\label{(33)}m\frac{d^2x}{dt^2}=-\nabla \left( V+Q\right) ,
\end{equation}
is obtained as subjected to the initial condition
\begin{equation}
\label{(34)}p=-\nabla S.
\end{equation}

Now assuming that $x$ represents a single particle's coordinate and $p$ it's
momentum, a spatial-temporal description similar to Newton's is obtained.

Yet this association of meanings cannot be made within the realm of the
present theory. In fat, the starting point was an equation that presents
dispersion relations according to the uncertainty principle; this dispersion
is included in the probability amplitude and shows up in the constituent
terms of (\ref{(28)}) and in equations (\ref{(29)}) and (\ref{(30)}). More
still, in the first article of this series it was demonstrated that
Schroedinger's Second Equation was not satisfied for s {\it %
ensembles}; for this reason equations (\ref{(29)}) and (\ref{(30)}) cannot
be associated to these {\it ensembles}. It is symptomatic that Bohm's theory
cannot be obtained from the equation for the density function, which is
dispersion free\cite{35}.

As was said in the first article, the equations satisfied by the individual
constituents of the systems are Newton's equations themselves. The equations
above represent no more than the representation of the {\it ensemble's}
behavior. The analysis of the double slit problem according to this
theory\cite{36} demonstrate this behavior clearly; it is also important to
note that the trajectories cannot be considered the real trajectories of the
systems' components, since we are using Schroedinger Second equation for the
probability amplitudes which is adequate only to describe the passage from
one equilibrium situation to another. According to equation (\ref{(34)}),
one cannot determine exactly the initial conditions of one of the {\it %
ensemble's} trajectory in particular, therefore leaving at least one hidden
variable which the formalism is not capable of revealing.

The fact that it is not possible to find a well defined physical source for
the quantum potential must also be stressed.

In this manner one concludes that Bohm's theory is not a hidden variable
theory and that the quantum potential should be interpreted as no more than
a fictitious ''potential'' representing a statistical field associated to a
particular problem's specific configuration.

It's non-local character is easily explained, once we have interpreted the
potential $Q\left( x\right) $ as a statistical potential. For Bohm's theory
for an {\it ensemble} of two particle systems, we have the following
equations
\begin{equation}
\label{(35)}\frac{d{\bf X}_1}{dt}=\rho \left( {\bf X}_1,{\bf X}_2\right)
^{-1}Im\sum_{ij}\Psi _{ij}^{*}\left( {\bf X}_1,{\bf X}_2\right) \frac
\partial {\partial {\bf X}_1}\Psi _{ij}\left( {\bf X}_1,{\bf X}_2\right) ,
\end{equation}
\begin{equation}
\label{(36)}\frac{d{\bf X}_2}{dt}=\rho \left( {\bf X}_1,{\bf X}_2\right)
^{-1}Im\sum_{ij}\Psi _{ij}^{*}\left( {\bf X}_1,{\bf X}_2\right) \frac
\partial {\partial {\bf X}_2}\Psi _{ij}\left( {\bf X}_1,{\bf X}_2\right) ,
\end{equation}
where
\begin{equation}
\label{(37)}\rho \left( {\bf X}_1,{\bf X}_2\right) =\sum_{ij}\left| \Psi
_{ij}\left( {\bf X}_1,{\bf X}_2\right) \right| ^2
\end{equation}
and ${\bf X}_1$ and ${\bf X}_2$ represent the particles' ''positions''\cite{%
37}. It is clear that, for such a system, every time that we cannot
write the density as a product
\begin{equation}
\label{(38)}\Psi _{ij}\left( {\bf X}_1,{\bf X}_2\right) =\Phi \left( {\bf X}%
_1\right) \Xi \left( {\bf X}_2\right) ,
\end{equation}
we will have the equations for ${\bf X}_1$ dependent upon ${\bf X}_2$ and
{\it vice-versa}, thus showing a non-local character.

As we have said, the potential $Q\left( x\right) $ represents the
statistical field associated to the {\it ensemble}. We should consider
equations (\ref{(35)}) and (\ref{(36)}) as representing only the property of
conditional probabilities. In other words, if we fix the statistical
behavior for one of the particles, we will know the statistical behavior of
the other\cite{38}.

\section{Conclusions}

In this series of papers we have presented a complete reconstruction of
quantum mechanic's principles. In this and the other papers, we demonstrated
that it is possible to interpret quantum from a Realist point of view. It
was also shown that the formalism itself obtained embraces that of usual
quantum mechanics as a particular one. Quantum mechanics was shown to be
local and, although statistical by principle, based on a deterministic
theory that is nothing more than newtonian classical mechanics. This
approach has also made possible for us to obtain a general relativistic
quantum theory for {\it ensembles} of systems with one particle.

All this collected, we are in position to maintain Einstein's definition of
reality\cite{17}. His affirmation about the incompleteness of quantum
mechanics, with its Complete Sets of Commuting Operators, that is, based on
the Schroe\-din\-ger equations for the probability amplitudes, can be
interpreted as a straightforward implication of the formalism connected with
Heisemberg's uncertainty relations.

Quantum mechanics, as represented by the first Schroedinger's equations, was
shown to be local and based on a deterministic Nature. Classical ontology
must be reconsidered\cite{39}.

\end{document}